# Deep Learning-enabled Spatial Phase Unwrapping for 3D Measurement

Xiaolong Luo[1], Wanzhong Song[1,*], Songlin Bai[1], Yu Li[2], and Zhihe Zhao[2]

[1] *College of Computer Science, Sichuan University, Chengdu 610065, China.*

[2] *State Key Laboratory of Oral Diseases, National Clinical Research Center for Oral Diseases, Department of Orthodontics, West China Hospital of Stomatology, Sichuan University, Chengdu, 610041, China.*

*\*songwz@scu.edu.cn*

**Abstract:** In terms of 3D imaging speed and system cost, the single-camera system projecting single-frequency patterns is the ideal option among all proposed Fringe Projection Profilometry (FPP) systems. This system necessitates a robust spatial phase unwrapping (SPU) algorithm. However, robust SPU remains a challenge in complex scenes. Quality-guided SPU algorithms need more efficient ways to identify the unreliable points in phase maps before unwrapping. End-to-end deep learning SPU methods face generality and interpretability problems. This paper proposes a hybrid method combining deep learning and traditional path-following for robust SPU in FPP. This hybrid SPU scheme demonstrates better robustness than traditional quality-guided SPU methods, better interpretability than end-to-end deep learning scheme, and generality on unseen data. Experiments on the real dataset of multiple illumination conditions and multiple FPP systems differing in image resolution, the number of fringes, fringe direction, and optics wavelength verify the effectiveness of the proposed method.

1. Introduction

Optical 3D shape measurement techniques have been applied in many fields such as manufacturing, entertainment, and healthcare [1]. Among optical 3D shape measurement techniques that have been proposed, the single-camera FPP system projecting only single-frequency patterns is ideal for 3D imaging speed and system cost. However, this system requires a robust spatial phase unwrapping (SPU) method for recovering the continuous phase from the wrapped phase map [2]. Many SPU methods have been presented, generally including path-dependent methods [3][4], path-independent methods [5][6][7], and deep learning-based methods [8][9][10][11][12][13][14][15]. This study focuses on the SPU in FPP, in which path-dependent methods are probably the most used methods; thus, path-independent methods are not discussed below.

Error propagation and phase discontinuity remain open challenges for path-dependent phase unwrapping methods. Quality-guided unwrapping is the primary strategy for dealing with error propagation in SPU [3][16]. This strategy defines a quality metric (also known as reliability) for each point of the phase map and uses this quality metric to guide the unwrapping path. Points of lower quality metrics are often regarded as unreliable ones with heavy noise. During the unwrapping, points of higher quality metrics are unwrapped earlier.

Many quality metrics have been suggested under the quality-guided SPU (QGSPU) framework over the decades, and their effectiveness has been demonstrated in various scenes. There exists a high correlation between phase errors and the fringe images' signal-to-noise ratio (SNR). In FPP, the intensity modulation in the areas of local shadow and low surface reflectivity is lower than that of other areas;

consequently, intensity modulation is a helpful guide to determining an optimal unwrapping path [17]. Data modulation is also suggested to reflect the SNR of fringe images [18]. Other quality maps are derived directly from the phase map, including second phase difference and amplitude map of transform-based methods. Second phase difference provides a measurement for the degree of concavity/convexity of the phase function, which can be utilized to detect possible phase inconsistencies between points [19][20][4]. Amplitude map of transform-based methods, such as Windowing Fourier Filtering (WFF) [21], shows good performance in identifying the unreliable points of phase maps. Zhao et al[16]. compared various QGSPU methods and found that transform-based methods show better performances under noisy conditions. And, QGSPU methods are hard to find a quality map that can best guide the phase unwrapping in the presence of phase discontinuity.

Most of the existing quality metrics for QGSPU are defined over a single data source, such as modulation, phase map, or amplitude map. Background intensity, modulation, and phase map all contain information that helps identify the unreliable points before unwrapping. However, traditional hand-crafted quality metrics in SPU lack effective ways to integrate the quality information from multiple sources, such as background intensity, modulation, and phase map. This lack of ways of integration limits the application scopes of SPU methods. In addition, manual adjustment of the quality threshold for correctly unwrapping is an indispensable step in many applications. This manual adjustment not only necessitates more tricks but also hinders the automation of 3D Measurement.

Besides QGSPU's strategy of detecting unreliable points before unwrapping, many post-unwrapping approaches have been proposed to detect unreliable points after unwrapping [22][23][24][25][26][27][28][29]. Post-unwrapping detection methods are not suitable for the continuous phase recovered by QGSPU but suitable for the continuous phase obtained with the temporal phase unwrapping (TPU). This study focuses on the SPU in FPP; more details of the post-unwrapping detection methods are not discussed here.

Deep learning SPU methods [8][9][10][11][12][13][14][15] demonstrated good denoising and unwrapping performance and outperformed the conventional path-dependent and path-independent methods. However, as discussed by [13], the training datasets used in many deep learning SPU unwrapping methods are simulated instead of real scenes. And many of the evaluation experiments were conducted on closed fringe patterns instead of the carrier fringe patterns common in FPP. For the phase unwrapping of carrier fringe patterns, Liang et al. [11]proposed a two-step strategy to unwrap single-frequency phase maps of real objects, and Bai et al. [15]demonstrated that recovering the absolute phase from a single-frequency high resolution (i.e., 1024×1024 pixels) phase map in real-time is feasible. These deep learning SPU methods are end-to-end learning, which means that their unwrapping performance is system and environment-dependent; the trained neural networks need to be fine-tuned before being applied to a different FPP system. Moreover, end-to-end deep learning unwrapping methods face the interpretability problem.

This study aims to improve the performance of SPU in FPP on complex scenes using deep learning and improve the generality of learning-based SPU methods. The main contributions are as follows,

(1) A method is presented for generating composite images for unreliable point detection before unwrapping in FPP. The method is suitable for normalizing the data distribution of scenes with different illumination conditions.

(2) Deep neural network is employed to learn the discriminative features of unreliable points from the composite images. This implementation efficiently integrates the information of reliable points from multiple maps and significantly improves the robustness of SPU in complex scenes.

(3) A two-stage hybrid SPU scheme is proposed. Compared with traditional QGSPU methods, this scheme avoids the manual adjustment of the masking threshold to adapt to the different scenes. In addition, this hybrid scheme has better generality on unseen data and interpretability than the end-to-end deep learning scheme.

The rest of this paper is organized as follows. Section 2 introduces the proposed method. Section 3 shows the experiment results. Section 4 discusses the proposed method's advantages, limitations, and future work. Section 5 outlines the conclusion.

## 2. Proposed method

Figure 1 depicts the proposed method's overall workflow. The structure of single-projector and single-camera FPP is employed. Wrapped phase, modulation, and background intensity are extracted from fringe images using phase-shifting technology to build one three-channel composite Phase-Modulation-Intensity (PMI) image. The PMI image is input into a lightweight deep convolutional neural network (DCNN), and the network outputs a pointwise classification map. A binary unwrapping mask is generated from the classification map to filter out unreliable points before unwrapping, and the filtered wrapped phase is unwrapping using the flood-fill algorithm.

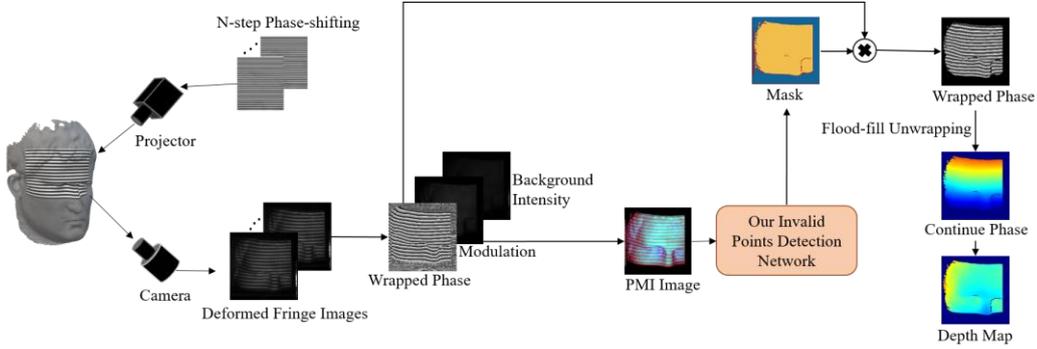

Fig. 1. Workflow of the proposed method. A light-weighted DCNN is trained to detect unreliable points from the single-frequency composite image. After the invalid points are identified, the complex phase map is unwrapped with the flood-fill algorithm.

*2.1 Phase-shifting profilometry*

Phase-shifting algorithms are widely used in 3D shape measurement due to their robustness and accuracy. A number of fringe images having the same wavelength with a certain phase shift are used to obtain the phase $\Phi$. For example, fringe images of an $N$-step phase-shifting algorithm with a phase shift of $2\pi/N$ can be represented as:

$$I_n(i,j) = I'(i,j) + I''(i,j)cos[\Phi(i,j) + 2n\pi/N], \quad (1)$$

$I_n(i,j)$ is the $k$th phase-shifted fringe image, $I'(i,j)$ indicates the background intensity, $I''(i,j)$ is the intensity modulation, $\Phi(i,j)$ represents the continuous phase containing geometric information of the object surface, $N$ is the number of phase-shifting steps, and $n = 0,1,...,N-1$. The wrapped phase $\varphi(i,j)$ can be retrieved as follows

$$\varphi(i,j) = arctan\left[\frac{\sum_{n=0}^{N-1} I_n(i,j)\sin(2n\pi/N)}{\sum_{n=0}^{N-1} I_n(i,j)\cos(2n\pi/N)}\right]. \quad (2)$$

Phase unwrapping can be interpreted as finding a unique integer number (i.e., fringe order) at each point and then eliminating $2\pi$ discontinuities by adding the integral multiple of $2\pi$ to the phase $\varphi(i,j)$.

The relationship between the wrapped $\varphi(i,j)$ and continuous $\Phi(i,j)$ phase is

$$\Phi(i,j) = \varphi(i,j) + 2\pi k(i,j), \tag{3}$$

here, $k(i,j)$ is the fringe order.

The background intensity $I'(i,j)$ and intensity modulation $I''(i,j)$ could also be recovered from the deformed fringe images,

$$I'(i,j) = \frac{1}{N}\sum_{n=0}^{N-1} I_n(i,j) \tag{4}$$

$$I''(i,j) = \frac{2}{N}\sqrt{[\sum_{n=0}^{N-1} I_n(i,j)sin\,(2n\pi/N)]^2 + [\sum_{n=0}^{N-1} I_n(i,j)cos\,(2n\pi/N)]^2} \tag{5}$$

For QGSPU, only one phase map is provided, the $k(i,j)$ of Eq. (3) is relative fringe order. In this study, the ground truth absolute fringe order for experimental comparisons is obtained using temporal phase unwrapping (TPU), the relative-to-absolute alignment of the fringe order of the comparative QGSPU methods, and our method is accomplished by making the fringe order difference of each subregion zero. But, in the 3D reconstruction experiment with our method, the absolute fringe order is obtained with the method by [30].

To obtain the 3D shape from the absolute phase $\Phi(i,j)$, we utilized the calibration method by [30]. For each arbitrary point $(x_c, y_c)$ on the camera's sensor plane, the absolute phase value is used to identify its corresponding point $(x_p, y_p)$ on the DMD plane according to the epipolar constraint as follows:

$$(x_p, y_p, 1)\boldsymbol{F}(x_c, y_c, 1)^T = 0, \tag{6}$$

here, $\boldsymbol{F}$ is the fundamental matrix. The 3D coordinate $(X, Y, Z)$ is calculated by the triangulation [31] as,

$$\begin{cases}(x_c, y_c, 1) = \boldsymbol{P}_c(X,Y,Z)^T \\ (x_p, y_p, 1) = \boldsymbol{P}_p(X,Y,Z)^T\end{cases} \tag{7}$$

here, $\boldsymbol{P}_c$ is the camera matrix and $\boldsymbol{P}_p$ is the projector matrix obtained via system calibration.

*2.2 Unreliable points removal before unwrapping*

For QGSPU in FPP, unreliable points should be identified before unwrapping. Unreliable points in the wrapped phase mainly include background points and unreliable object points. Background points could be removed through modulation thresholding. Unreliable object points often appear in the areas of low surface reflectivity, shadow, motion blur, and phase discontinuity. For simple scenes, the modulation of reliable points is often larger than that of unreliable ones, and points near discontinuous regions are often reliable with high probability. Using such cues for unreliable object points detection is straightforward and traditional algorithms such as thresholding on modulation, background intensity, and phase gradient before unwrapping can easily detect these unreliable points without training. However, even for simple scenes, different scenes need different thresholds for correctly unwrapping, and selecting a proper threshold requires many trials. A larger threshold often ensures correct unwrapping at the cost of filtering more reliable points.

For complex scenes, such as low surface reflectivity, motion blur, and discontinuity exist at the same time, the points adjacent to phase discontinuities and the points in shadow areas, low surface reflectivity areas, and motion blur areas are unreliable points, thresholds for correctly unwrapping are troublesome to successful 3D reconstruction. Determining how to integrate these thresholds becomes difficult and needs more trials with tricks. A common result for complex scenes is removing excess reliable points to ensure correct unwrapping when using SPU. This can be observed in Fig. 2, which illustrates one example of phase unwrapping by three traditional QGSPU methods in complex scenes. The three QGSPU method are MODU-sort [17], FSPU [19], and WFF-sort [21]. The ground truth of the continuous phase with only reliable points is obtained using TPU and the outliers detection method discussed in Section 2.3. It can be observed that all the three traditional QGSPU methods delete excess object points in order to unwrap correctly. Although background intensity, modulation, and phase map all contain information that helps identify the unreliable points before unwrapping, QGSPU methods lack effective ways to integrate quality information from multiple sources. This makes QGSPU methods face challenges when dealing with the complex phase map.

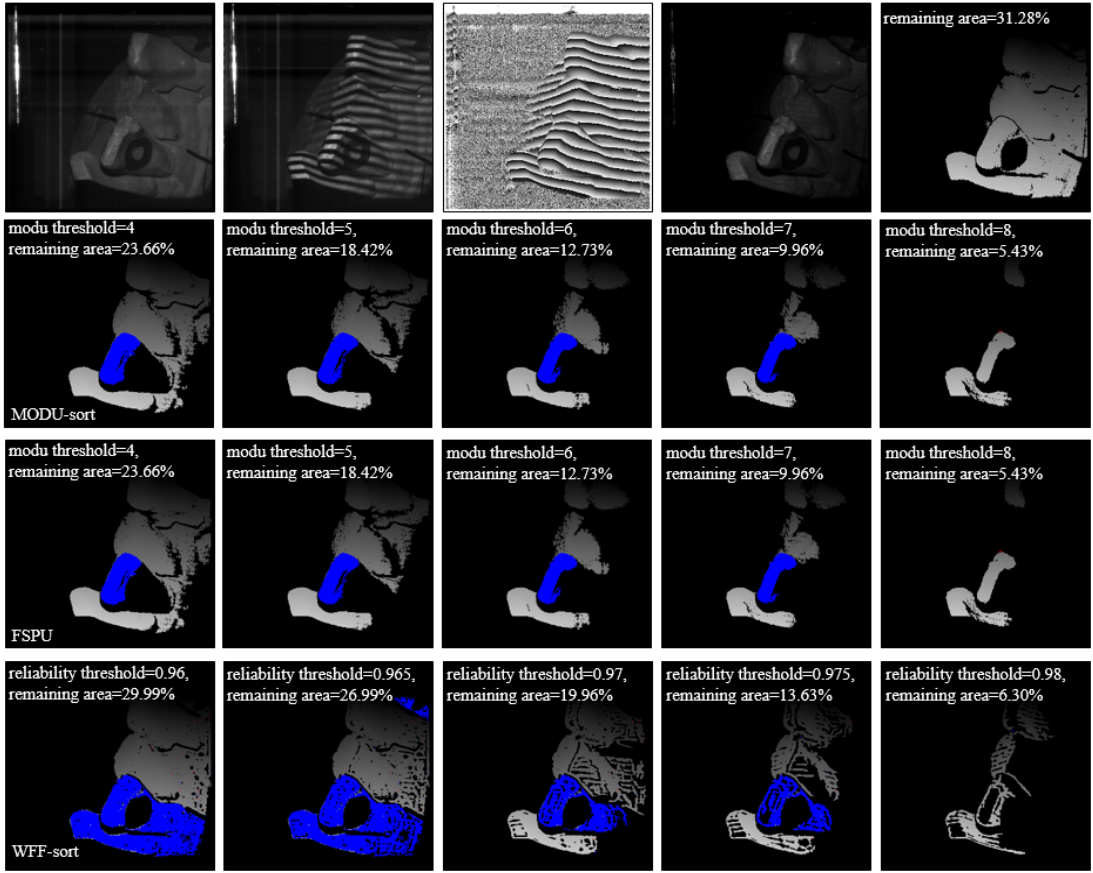

Fig. 2. QGSPU methods remove excess reliable points to ensure correctly unwrapping. The first row is amplified background intensity, amplified fringe image, wrapped phase, amplified intensity modulation, and the ground-truth continuous phase without unreliable points, from left to right. The second to fourth row shows the continuous phase recovered by different methods at different thresholds, and blue points show the points whose continuous phase is 0.3 rad lower than their ground truth.

Deep neural networks have obvious advantages in learning discriminative features from multiple images. On the basis of these learned features, neural networks could classify each point of phase maps as reliable or unreliable. Under the framework of pixel-wise classification, one possible way of implementing unreliable point detection in phase maps using DCNN is to output a two-class classification map, in which each point is classified as reliable or unreliable. However, in this two-class classification scheme, background points and unreliable object points belong to the same class. For scenes where the proportion of background points is significantly higher than that of unreliable object points, the neural network will be prone to learn how to identify the background points instead of the unreliable object points.

This study adopts the three-class classification scheme. The three classes of points include background points, unreliable object points, and reliable points. Phase map, modulation, and background intensity are combined into one three-channel image (i.e., PMI image), which is taken as the input by one deep neural network. After training, the deep neural network outputs a classification map for each PMI image. With this classification map, a binary mask is built. In this mask, only the points classified as reliable are unwrapped by the flood-fill algorithm.

*2.3 Training dataset*

We used the same handheld FPP system in [15] to collect the data for training the neural network. This FPP system consists of one projector unit of DLP LightCrafter with 684×608 DMD pixels and one CMOS camera with a resolution of 1024×1024 pixels; its measurement volume is approximately 16 mm×16 mm×8 mm. We employ phase-shifting binary patterns to generate the wrapped phase for obtaining better image contrast in an oral environment. Binary patterns require more steps of phase-shifting than sinusoidal patterns to produce a wrapped phase of comparable quality. On the other hand, handheld FPP systems have a minimum requirement for a 3D scan-reconstruction frame rate (i.e., ten fps). This requirement limits the step number of phase-shifting and the fringe frequency. Considering these factors, we used the complementary Gray code [32] and phase-shifting of binary patterns to collect the training data and generate the ground truth of continuous phase and depth map. The period of binary pattern 42, the number of phase-shifting steps is 14, and the number of fringes is 16.

Input images in the training dataset are generated from the fringe images through the following five-step procedure. First, wrapped phase, background intensity, and modulation were retrieved using Eq. (2), Eq. (4), and Eq. (5), respectively. Second, a modulation threshold of two was set to clean background points from the three maps. Third, four-connected neighborhood detecting was performed on three maps, and any small region whose number of points is less than one percent of the total number of points is deleted. Fourth, an intra-frame normalization method was conducted on the modulation and background intensity. Finally, the phase map was normalized to [0,1] and stacked with the normalized modulation and the normalized background intensity into one three-channel PMI image.

Different FPP systems often work under various illumination conditions. Even for better generality, feeding DCNNs with a large number of data from various FPP systems is undesirable and laborious. We propose the intra-frame normalization method to normalize the different distributions caused by the various illuminations. This intra-frame normalization works as follows, all values of one map are sorted descendingly, and then the top one percent of points are removed. After that, the map is linearly normalized as

$$\bar{I}(i,j) = \begin{cases} 1, & I(i,j) > T_{\max} \\ \frac{I(i,j)}{T_{max}}, & otherwise \end{cases}, \quad (8)$$

here, $I(i,j)$ is the filtered modulation or background intensity with a modulation threshold of two, $\bar{I}(i,j)$ is the corresponding normalized images, and $T_{\max}$ is the highest value on the remaining queue.

The training and validation datasets are collected using the same handheld FPP system under the same illumination conditions. The training dataset contains 8,087 images of 40 dental casts; the validation dataset includes 2,451 images of two dental casts. Figure 3 shows examples of the acquired fringe images and the training dataset. All images have a resolution of 1024×1024 pixels. Although the global shapes of dental casts are similar, local shapes within the measurement volume (i.e., 16 mm×16 mm×8 mm) of the FPP system are significantly distinctive. Various local shape changes ensure a variety of surface shapes in the training data.

Please note that the camera of the handheld FPP system is custom-designed instead of a commercial off-the-shelf product. Fringe images are converted from the RAW data of the CMOS sensor. During the transformation from RAW data to images, except for a fixed gain parameter and automatic black level, no other image signal processing tasks, such as exposure correction, denoising, sharpening, and gamma correction, were performed. Therefore, the intensity value of the fringe images from our FPP system is relatively low, as shown in the second column of Fig. 3.

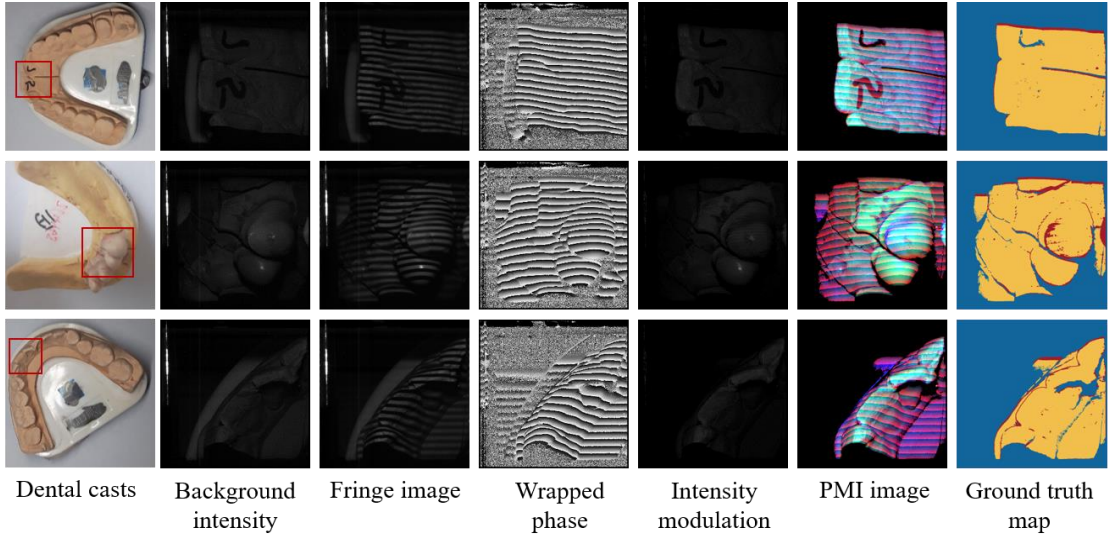

Dental casts | Background intensity | Fringe image | Wrapped phase | Intensity modulation | PMI image | Ground truth map

Fig. 3. Examples of the training data collected under the same illumination. The first column shows three dental casts of the training data. The second column to the fifth column is background intensity, fringe image, wrapped phase, and modulation. The sixth column is the composite PMI image taken as the input by the DCNN. The seventh column is the ground-truth classification map. All images have a resolution of 1024×1024 pixels.

In supervised learning, each point of the PMI image needs to be labeled as one of the three classes. Labeling of unreliable object points is accomplished by detecting unreliable points from the continuous phase recovered with TPU algorithms. We adopted a registration-orientation automatic labeling strategy. This labeling strategy assumes that unreliable object points in phase maps correspond to the unreliable points in depth maps. Unreliable points in depth maps will become outliers during the registration and fusion of point clouds. Thus, labeling the unreliable object points in phase maps is equivalent to detecting outliers in point clouds. In our handheld FPP system, outliers in point clouds are detected through an automatic pipelined procedure, which includes bilateral filtering, local depth-variance-maximum suppression, and small region removal. Figure 4 shows one example of outliers detection results. After this process, the depth value of the unreliable points in the depth map was set to zero. Finally, the filtered depth map and the modulation are used to create the labels of each point as

$$class = \begin{cases} 0, modu \leq 2 \\ 1, modu > 2 \text{ and } depth = 0, \\ 2, otherwise \end{cases} \quad (9)$$

here, class-0 represents background points, class-1 denotes unreliable points, and class-2 represents reliable points.

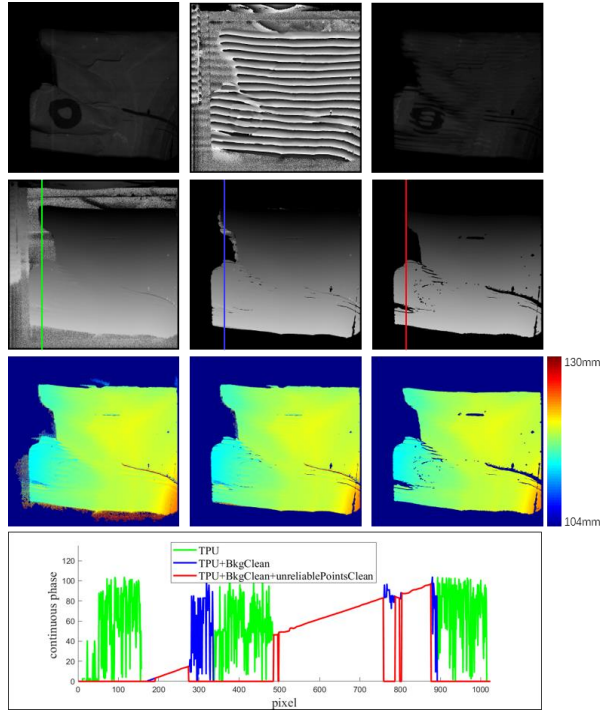

Fig. 4. Labeling of unreliable points using results of outliers detection during point clouds registration. The first row is background intensity (left), wrapped phase, and intensity modulation. The second row shows the original continuous phase from TPU, the continuous phase without background points, and the continuous phase without background points and unreliable object points. The third row shows the depth maps corresponding to the three continuous phase maps in the second row. The fourth row is the continuous phase profiles.

*2.4 Network architecture*

With the increasing use of high-resolution (i.e., 1024×1024 pixels or greater) fringe images in FPP applications, learning-based fringe analysis requires neural networks to be able to process high-resolution fringe images. In this study, a lightweight network ERFNet [33], was chosen as the baseline. Different from images with which the general semantic segmentation task deal, for the phase maps of FPP, the regions belonging to the same class (i.e., fringe order) are less varied in texture, and regions belonging to different fringe orders have similar textures. We replaced the max-pooling operator in each down-sampling layer of ERFNET with a mean-pooling operator, as by [15] , to retain more information on poor-texture regions.

The network structure used in this study is shown in Fig. 5. The size of input images is $1024 \times 1024 \times 3$ pixel, ReLU activation function and batch normalization (BN) are used in each layer, and softmax was used in the last layer to normalize the predicted probabilities before calculating the loss. The loss function is the standard weighted cross-entropy. The detailed information for each layer is listed in Table 1.

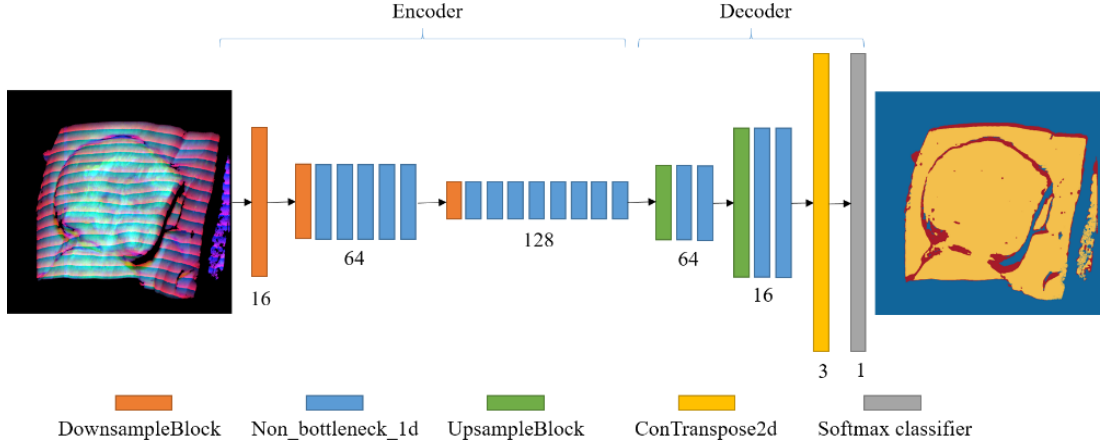

Fig. 5. Structure diagram of the invalid points detection network. (In the output map, blue, red, and yellow indicate background points, unreliable points, and reliable points, respectively.)

Table 1. Layer disposal of our unreliable detection network. The input size is 1024×1024×3.

|  | Layer | Output size |
|---|---|---|
|  | DownsampleBlock | $512 \times 512 \times 16$ |
|  | DownsampleBlock | $256 \times 256 \times 64$ |
|  | 5×Non_bottleneck_1d | $256 \times 256 \times 64$ |
|  | DownsampleBlock | $128 \times 128 \times 128$ |
|  | Non_bottleneck_1d(dilated 2) | $128 \times 128 \times 128$ |
|  | Non_bottleneck_1d(dilated 4) | $128 \times 128 \times 128$ |
| Encoder | Non_bottleneck_1d(dilated 6) | $128 \times 128 \times 128$ |
|  | Non_bottleneck_1d(dilated 8) | $128 \times 128 \times 128$ |
|  | Non_bottleneck_1d(dilated 2) | $128 \times 128 \times 128$ |
|  | Non_bottleneck_1d(dilated 4) | $128 \times 128 \times 128$ |
|  | Non_bottleneck_1d(dilated 6) | $128 \times 128 \times 128$ |
|  | Non_bottleneck_1d(dilated 8) | $128 \times 128 \times 128$ |
|  | UpsampleBlock | $256 \times 256 \times 64$ |
|  | 2×Non_bottleneck_1d | $256 \times 256 \times 64$ |
| Decoder | UpsampleBlock | $512 \times 512 \times 16$ |
|  | 2×Non_bottleneck_1d | $512 \times 512 \times 16$ |
|  | ConvTranspose2d | $1024 \times 1024 \times 3$ |
|  | Softmax classifier | $1024 \times 1024 \times 3$ |

## 3. Experiments

Our unreliable points detection network is implemented using Pytorch 1.8.0, and trained on one NVIDIA Titan RTX (CPU i7-7820X, 3.6 GHz, and 32 GB DDR4 RAM) with CUDA 10.2. Adam optimizer with a momentum of 0.9, an initial learning rate of 0.0005, a dropout ratio of 0.03, and 0.3 in the encoder was set. The number of epochs is 100, and the batch size is 16. To further improve the generality of the trained network, random data augmentation, including horizontal-flip, vertical-flip, and rotation of random angles, were employed. Our unreliable points detection network is trained only with the data collected under an operating voltage of 4.99V. The start point of the flood-fill algorithm is the midpoint of the one-dimensional sequence column-orderly flattened from the region. It is worth noting that we trained on the full-resolution (1024×1024) pixels image rather than the down-sampled image.

Three QGSPU methods, including MODU-sort [17], FSPU [19], and WFF-sort [21], and one deep learning SPU method, i.e., Hi-Phase [15], were selected as the comparative methods. For MODU-sort and FSPU, we used the implementation of I2L2 by [34] and the implementation by [35]. For WFF-sort, the implementation by [21][34] is used. For Hi-Phase, the pre-trained network [36] was directly used without retraining.

Our method and the three QGSPU methods only retrieve the relative continuous phase, aligning the relative phase to the absolute phase through adding/subtracting a global integer multiplies of $2\pi$ for each subregion is conducted before comparing the results of different methods. This global integer is determined by selecting the fringe-order difference corresponding to the subregion. In addition, all comparisons between a specified SPU method and the ground truth were conducted over the valid points.

*3.1 Test datasets*

Test data from four different FPP systems were collected to evaluate the proposed method's effectiveness and generality. All test data have an absolute phase recovered by TPU. As illustrated in Table 2, these systems differ in optics wavelengths, measurement areas, numbers of fringes, fringe direction, or type of fringe pattern. Some of them also have different image resolutions or image contrast. The training and validation datasets are generated from defocused binary pattern images, and the test data are generated from binary and sinusoidal pattern images. The characteristics of the four FPP systems are noted below.

Table 2. Four FPP systems with different characteristics were used to collect the test data. The training dataset is collected using System-A under one illumination condition.

| | System-A (Test-Data-A) | System-B (Test-Data-B) | System-C (Test-Data-C) | System-D (Test-Data-D) |
|---|---|---|---|---|
| Light engine | DLP | laser MEMS | DLP | DLP |
| Optics wavelength | 450nm | 820nm | 730nm | white light |
| Measurement area (mm$^2$) | 16×16 | 400×500 | 400×500 | 400×500 |
| Image resolution (pixel) | 1024×1024 | 960×1280 | 1440×1080 | 1024×1024 |
| Type of fringe pattern | binary | binary | sinusoidal | sinusoidal |
| Number of fringes | 16 | 64 | 49 | 42 |
| Direction of fringe | horizontal | vertical | vertical | horizontal |
| Multiple illuminations | Y | N | N | N |
| Train dataset | Y | N | N | N |
| Test dataset | Y | Y | Y | Y |

*Test-Data-A*

This dataset contains 5,632 maps captured under five illumination conditions with the handheld FPP system (referred to as System-A) described in section 2.2. Ablation study, evaluation metric selection, comparative experiments, and 3D reconstruction experiments are conducted using this dataset. Because the light engine of System-A is developed by ourselves, we can change the operating voltage of the LED light source to achieve different illumination intensities of the fringe projection. One dental cast is scanned under multiple illuminations to build Test-data-A; This dental cast differs from the dental models of the training and validation datasets. Figure 6 shows examples of Test-Data-A. Please note that both the training dataset and the validation dataset are collected under the same illumination conditions (i.e., an operating voltage of 4.99V), and Test-Data-A is collected under five multiple illuminations, as listed in Table 3.

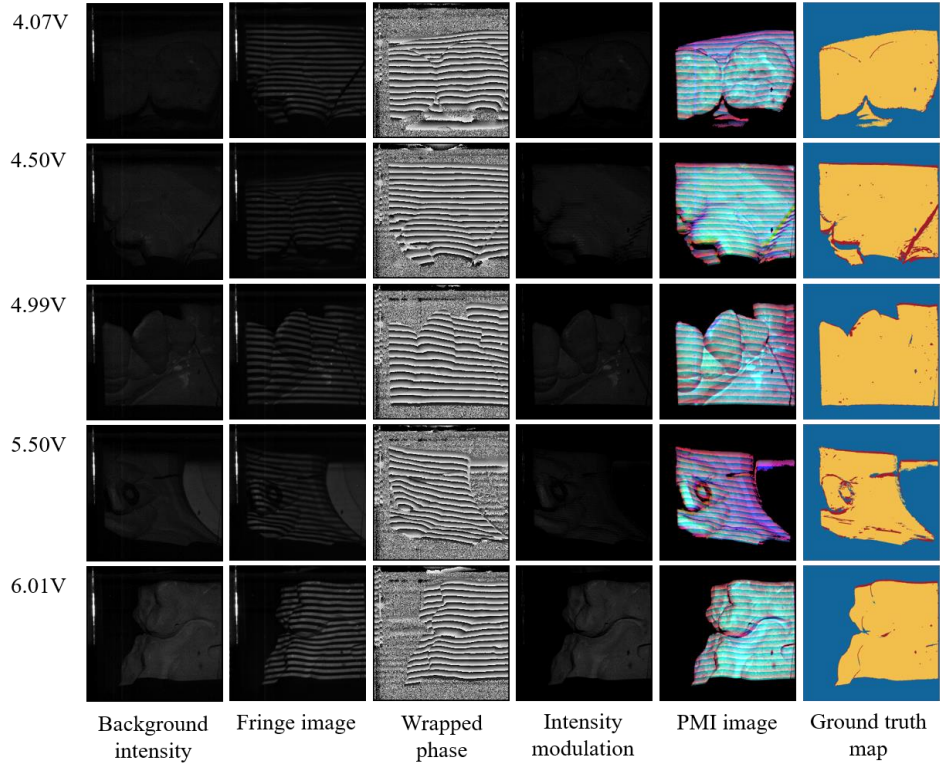

Fig. 6. Examples of Test-Data-A collected using System-A under five different illuminations. The first column is the background intensity. The second to fourth columns show the fringe image, wrapped phase, and modulation. The fifth column is the composite PMI image. The ground truth classification map is shown in the last column.

Table 3. Number of phase maps collected under five different illuminations in Test-Data-A

| Operating voltage (V) | 4.07 | 4.50 | 4.99 | 5.50 | 6.01 |
|---|---|---|---|---|---|
| Number of phase maps | 1,145 | 1,187 | 1,064 | 1,203 | 1,033 |

Test-Data-A covers phase maps of multiple typical types, including low surface reflectivity, motion blur, and phase discontinuity. Table 4 lists the statistics of types of phase maps in Test-Data-A.

Table 4. Number of different types of phase maps in Test-Data-A

| Operating voltage (V) | Simple | Low surface reflectivity or motion blur | Phase discontinuity | Complex |
|---|---|---|---|---|
| 4.07 | 149 | 288 | 161 | 547 |
| 4.50 | 109 | 259 | 222 | 597 |
| 4.99 | 74 | 252 | 151 | 587 |
| 5.50 | 104 | 147 | 195 | 757 |
| 6.01 | 78 | 191 | 161 | 603 |

*Test-Data-B*

Test-Data-B is collected with one UTECH infrared laser FPP system (referred to as System-B). This system consists of a laser Micro-Electro-Mechanical System (MEMS) projector working on an optics wavelength of 820nm with two black-white cameras and an RGB camera. Its measurement area is approximately 400 mm × 500 mm. It achieves 3D Measurement through three-frequency 4-step phase-shifting of binary patterns. The highest number of fringes is 64. The resolution of fringe images is 960×1280 pixels.

*Test-Data-C*

Test-Data-C is collected with one infrared DLP FPP system we built (referred to as System-C). This system consists of a GVINDA PDC03 light engine Kit working on an optics wavelength of 730nm and one infrared CMOS camera of MV-CA016-10UM. Its measurement area is approximately 400 mm × 500 mm. It achieves 3D Measurement through the three-frequency 4-step phase-shifting of sinusoidal patterns. The highest number of fringes is 49. The resolution of fringe images is 1440×1080 pixels.

*Test-Data-D*

Test-Data-D is collected from one WISESOFT white light DLP FPP system (referred to as System-D) with an image resolution of 1024×1024 pixels. Its measurement area is approximately 400 mm × 500 mm. It achieves 3D Measurement through the three-frequency 4-step phase-shifting of sinusoidal patterns. The highest number of fringes is 42. The resolution of fringe images is 1024×1024 pixels.

*3.2 Preprocessing*

Different SPU methods usually adopt different preprocessing to maximize their performance. The fringe images acquired under different illuminations have different modulation distributions, as shown in Fig. 7. For MODU-sort and FSPU, modulation thresholds of 6, 7, 8, 9, and 10 were set for the five different illuminations, respectively. All phase maps are first filtered using this modulation threshold to remove background points and some unreliable object points. This differentiated setting is to make Modu-sort and FSPU perform better under different scenes.

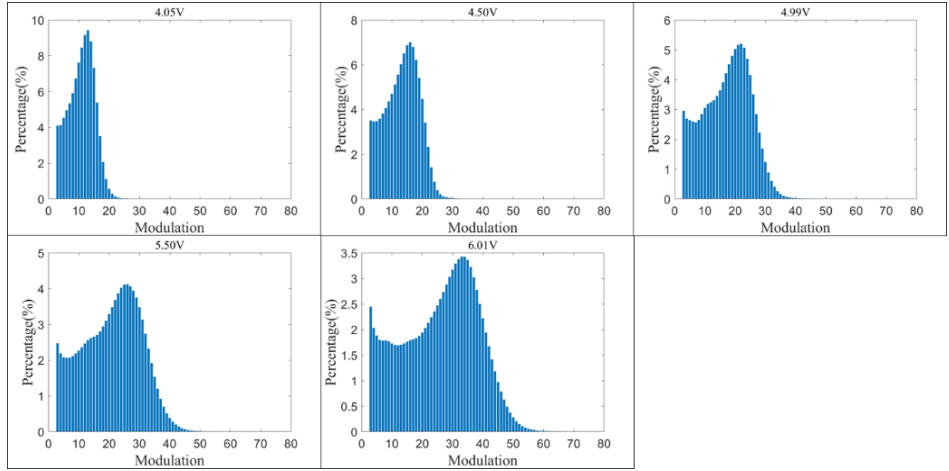

Fig. 7. Modulation distribution of the fringe images in Test-Data-A captured under five illuminations.

For WFF-sort implementation, we empirically configured it with a sigma of eight and other default parameters. And a threshold of 0.967 is set for the quality map of WFF-sort to remove background points and some unreliable object points. Modulation thresholding is not applied to WFF-sort because proportionally equivalent background points and unreliable object points have been deleted with the quality map threshold of 0.967.

For our method and HiPhase, a restrictive modulation threshold of two is used for all five illuminations. And this threshold is used for all the test data. This fixed setting can help to test and analysis the robustness of our method.

After the thresholding, four-connected neighborhood detection is performed, and any small region whose number of points is less than one percent of the total number of points is deleted. Then the phase maps are input into different methods for unwrapping. Figure 8 illustrates the average percentage of the remaining object points over all 5,632 phase maps in Test-Data-A after the preprocessing. The average number of remaining points in the input phase maps of the three QGSPU methods is greater than that of our method. This difference is due to different modulation thresholds set for scenes of different illuminations. This differentiated set is to make Modu-sort and FSPU perform better under different scenes.

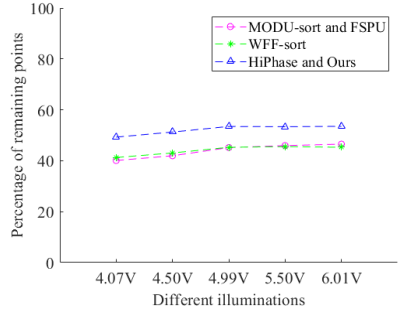

Fig. 8. Average percentage of remaining points of all phase maps in Test-Data-A after preprocessing.

*3.3 Evaluation metrics*

This study formulated the problem of unreliable-point detection from phase maps before unwrapping as a three-class pointwise classification problem. In general pointwise classification tasks, mean Intersection over Union (mIoU) is the most common evaluation metric. Frequency weighted intersection over union (FWIoU), mean pixel accuracy (mPA), pixel accuracy (PA), and class pixel accuracy (CPA) are also common metrics. Considering that our task is a handheld 3D measurement using FPP, we selected three metrics, including *number of failure cases* and *average RMSE of the depth map*, as candidates for evaluation metrics.

Except for WFF-sort, the continuous phase is regarded as a failure case when there is at least one region where the recovered order of all points differs from the ground truth and the number of points in this subregion is more than 0.1% (referred to as error-percent threshold) of the whole map. For WFF-sort, because it outputs smoothed continuous phase, the continuous phase is regarded as a failure case when there is at least one region where the recovered phase values of all points differ from the ground truth by more than 0.3 rad, and the number of points in this subregion is more than 0.1% of the whole map. The value of 0.3 rad is set empirically.

To investigate the effectiveness of the metrics mentioned above, we trained the original ERFNET, the network in-training is saved during each epoch of the training process, and each evaluation metric is calculated. The best epoch number for each metric is determined according to the highest score. The saved neural networks at the best epoch number corresponding to different metrics are tested on the 1,064 phase maps of 4.99V in Test-Data-A. Figure 9 shows the training and validation losses during the training process.

Table 5 illustrates the statistics of evaluation metrics. The RMSE is average over all 1,064 phase maps, including the failures (hereafter, the same calculation approach of RMSE). It can be observed the model with the highest value of CPA-1, i.e., CPA of the class of unreliable points, has the lowest number of failure cases and average RMSE of the depth map. In the following experiments, we trained the neural networks with a fixed number of epochs (i.e., 100) and selected the model with the highest CPA-1 value as the best model of our method for evaluating and comparing.

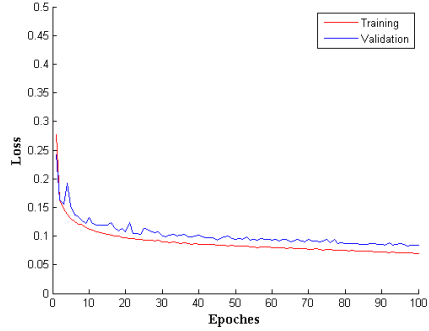

Fig. 9. Training and validation loss of the evaluation metrics selection experiment.

**Table 5. Statistics of different evaluation metrics averaged over 1,064 phase maps.**

|  | PA | mIoU | FWIOU | mPA | CPA-1* |
|---|---|---|---|---|---|
| Epoch number with the highest score | 92 | 92 | 92 | 100 | 80 |
| Number of failures | 90 | 90 | 90 | 74 | 60 |
| Average RMSE of the depth map (mm) | 0.080 | 0.080 | 0.080 | 0.064 | 0.042 |

*CPA-1 stands for CPA of class-1

### 3.4 Ablation study

Three ablation experiments were conducted to evaluate the effectiveness of the PMI composite images, intra-frame normalization, and mean-pooling replacement in the down-sampling block.

*Different channel combinations of the input composite image*

We trained multiple networks of ERFNET using the data of 4.99V with different input channel combinations. The number of failure cases and average RMSE of depth maps are listed in Table 6. The last two combinations of Table 6 have similar performance and outperform the other five combinations.

**Table 6. Average evaluation metrics of different channel combinations.**

| Different combinations as input | Number of failure cases | Average RMSE of the depth map (mm) |
|---|---|---|
| Background Intensity | 157 | 0.125 |
| Modulation | 137 | 0.095 |
| Phase | 122 | 0.133 |
| Phase + Modulation (PM) | 91 | 0.059 |
| Phase + Background Intensity (PI) | 71 | 0.060 |
| Modulation + Background Intensity (MI) | 69 | 0.039 |
| Phase + Modulation + Background Intensity (PMI) | 60 | 0.042 |

*Replacement of max-pooling with mean-pooling*

As discussed in Section 2.4, we replace the max-pooling operator with mean-pooling to help the neural network learn features from the poor-texture regions of phase maps. After this replacement, the data of the last combinations in Table 6 are used to train the neural network. Table 6 lists the average evaluation metrics after mean-pooling replacement. It can be seen that the combination of PMI outperforms the combination of MI when mean-pooling replacement is employed by comparing Table 6 and Table 7.

Table 7. Average evaluation metrics of mean-pooling replacement on 1,064 phase maps.

| Different combinations as input | Number of failure cases | Average RMSE of the depth map (mm) |
|---|---|---|
| MI | 89 | 0.056 |
| PMI | 44 | 0.027 |

*Intra-frame normalization*

To investigate the effectiveness of intra-frame normalization, we used the intra-frame normalized PMI images to train the neural network with mean-pooling replacement (referred to as mean-ERFNET). The training data is from the data collected at the operating voltage of 4.99V, and all the maps of Test-Data-A from five different illuminations are used as the test data. Table 8 shows the ablation results.

Table 8. Average evaluation metrics of intra-frame normalization on 5,632 phase maps of five different illuminations.

| The input of the neural network | Number of failure cases | Average RMSE of the depth map (mm) |
|---|---|---|
| PMI without IFN | 261 | 0.034 |
| PMI with IFN* | 226 | 0.033 |

* IFN means intra-frame normalization.

On the basis of the results of the above ablation study, we determined *number of failure cases* as the primary evaluation metric.

*3.5 Comparison in different scenes*

Figures 10-11 show the percentage of failure cases of different methods on Test-Data-A. In general, the failure percentages of the three QGSPU methods and our method increase with the increase of illumination. The primary reason is that there is a higher percentage of complex phase maps in the data collected at higher operating voltage, as shown in Table 3 and Table 4.

Figure 11 shows that phase discontinuity and complex scenes are challenging to the three QGSPU methods. Deep learning-based SPU methods show significantly better performance than traditional QGSPU methods. HiPhase's performance decreases in low and high illumination scenes due to the limited generality of the end-to-end SPU network on unseen data. Our method achieved a number of failure cases half that of HiPhase.

Figure 12 shows the average percentage of data points in phase maps unwrapped correctly with five methods in Test-Data-A. The number of correct data points from by HiPhase is closest to the ground truth. Under a total of five different illumination conditions, the number of correct data points output by our method is almost the same as that of the three QGSPU methods under two illumination conditions and slightly less than that of the three QGSPU methods under the other three illumination conditions.

There are 514 simple phase maps in Test-Data-A. Our method can unwrap them all. The number of failure cases of MODU-sort, FSPU, WFF-sort, and HiPhase on these simple phase maps is 4, 9, 10, and 49, respectively. Phase errors due to noise are the primary cause of failures of QGSPU methods. Limited generality is the primary reason for HiPhase's failure.

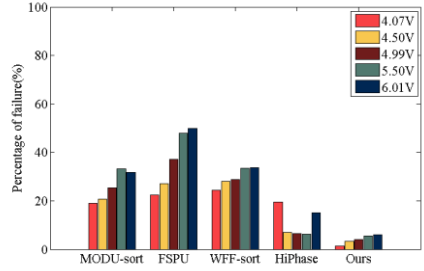

Fig. 10. Average failure percentage of five methods on the data of different illuminations in Test-Data-A.

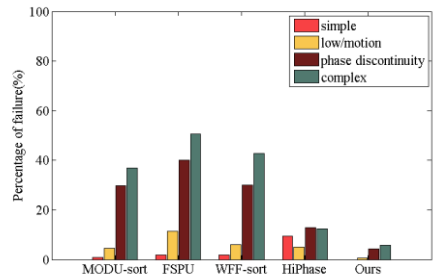

Fig. 11. Average failure percentage of five methods on the data of different types of phase maps in Test-Data-A.

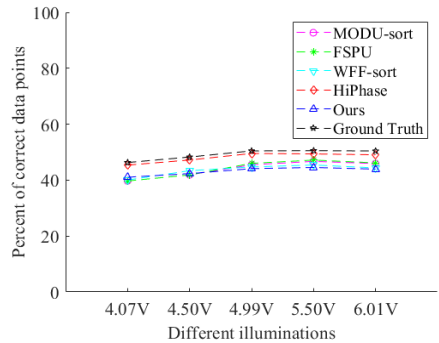

Fig. 12. Average percentage of data points in phase maps unwrapped correctly in Test-Data-A.

There are 1,137 phase maps with only surface reflectivity or motion blur. Low reflectivity and shadows are common in FPP systems. Motion blur causes phase errors in the phase map, especially in areas where the shape changes sharply. Under this condition, the number of failure cases of MODU-sort, FSPU, WFF-sort, HiPhase, and our method is 51, 129, 68, 56, and 7, respectively. Figures 13-14 illustrate the unwrapping results of the five methods considering low surface reflectivity and motion blur. The performance difference between our method and HiPhase shows the hybrid scheme works better than the end-to-end deep learning scheme in the presence of surface reflectivity or motion blur.

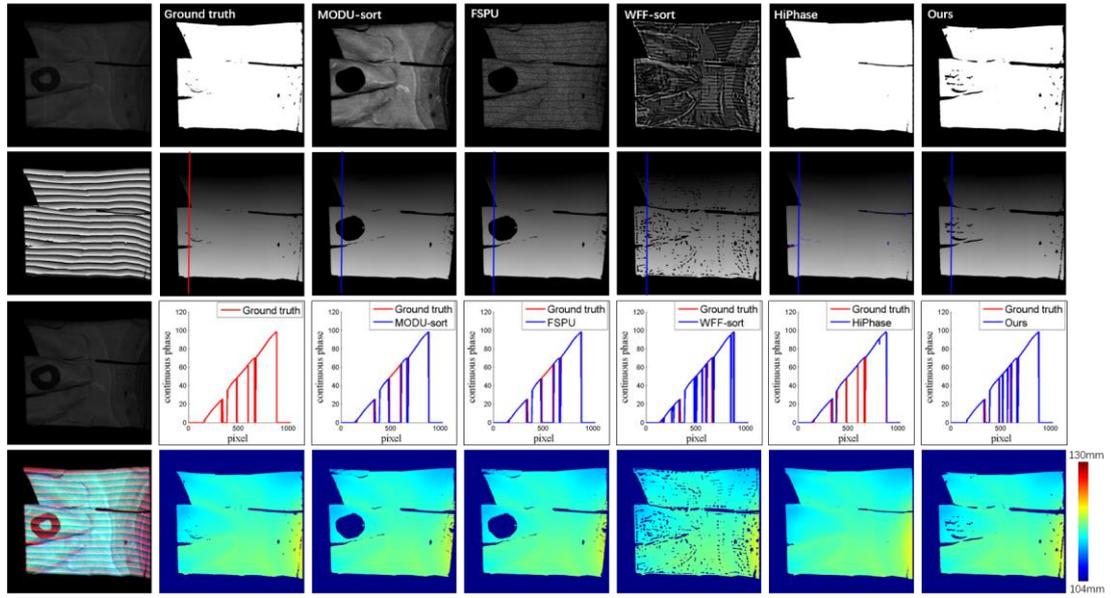

Fig. 13. Unwrapping results of five methods in low surface reflectivity scene. The first column shows the background intensity, phase map, intensity modulation, and PMI image from top to bottom. The second column is the ground truth. Results of the five methods are shown in the third to the seventh column; from top to bottom, the quality-map/unwrapping-mask, continuous phase, phase profile, and depth map are illustrated. For convenience, the value of the quality-map is linearly normalized to the interval [0, 255] and shown in a gray-level image.

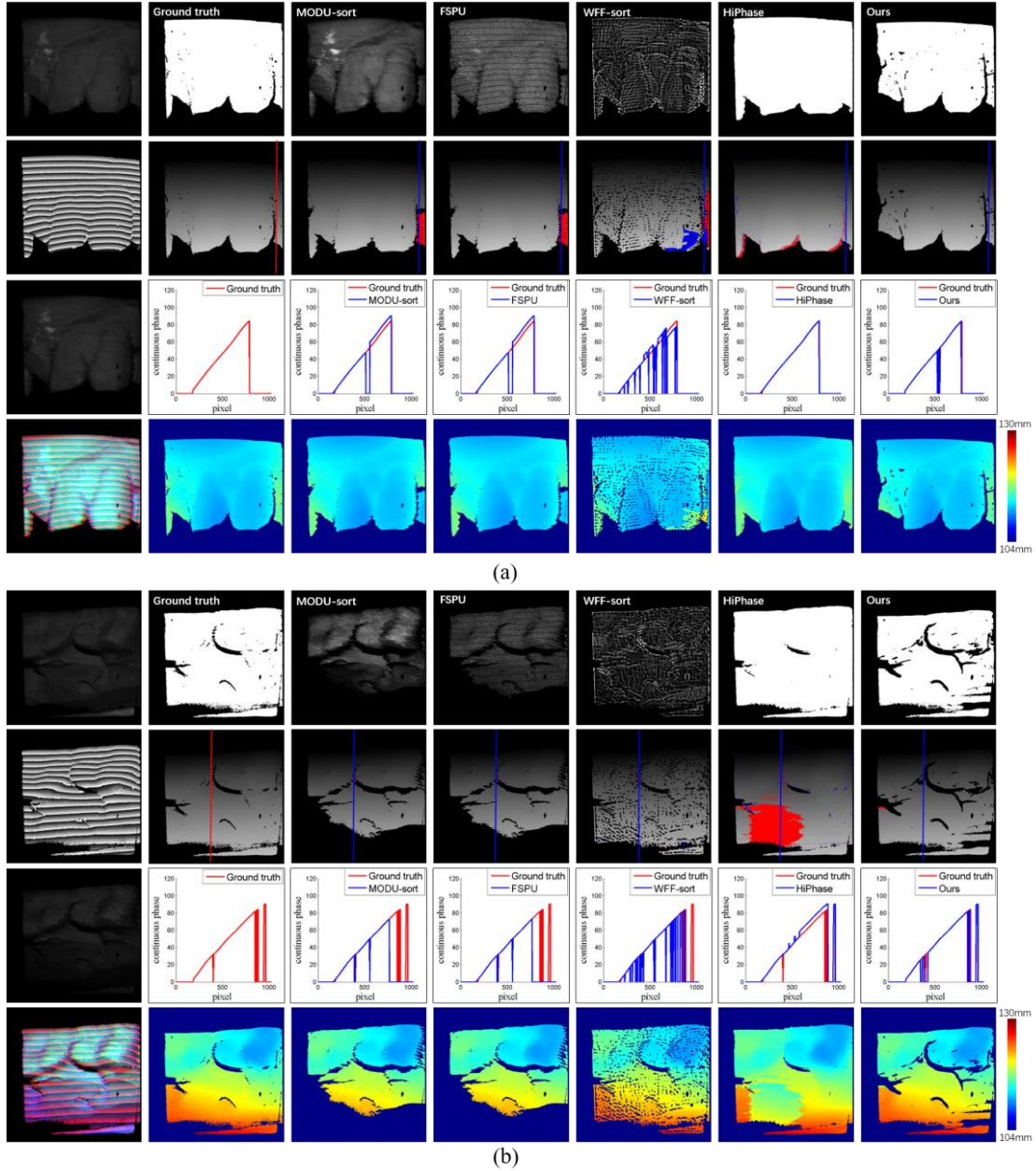

Fig. 14. Unwrapping results of five methods in motion blur scenes. In each group, the first column shows the background intensity, phase map, intensity modulation, and PMI image from top to bottom. The second column is the ground truth. Results of the five methods are shown in the third to the seventh column; from top to bottom, the quality-map/unwrapping-mask, continuous phase, phase profile, and depth map are illustrated. Red points in the continuous phase map indicate the points whose continuous phase is 0.3 rad bigger than their ground truth.

There are 890 phase maps with phase discontinuity. The number of failures of MODU-sort, FSPU, WFF-sort, HiPhase, and our method is 265, 356, 267, 115, and 39, respectively. Unwrapping results of the five methods in phase discontinuity scenes are shown in Fig. 15. In the presence of phase discontinuity, our method over-performs the other four methods. Another deep learning-based SPU method, HiPhase, achieved half the number of failure cases of traditional QGSPU methods. The performance difference between our method and HiPhase shows the hybrid scheme may work better than the end-to-end deep learning scheme.

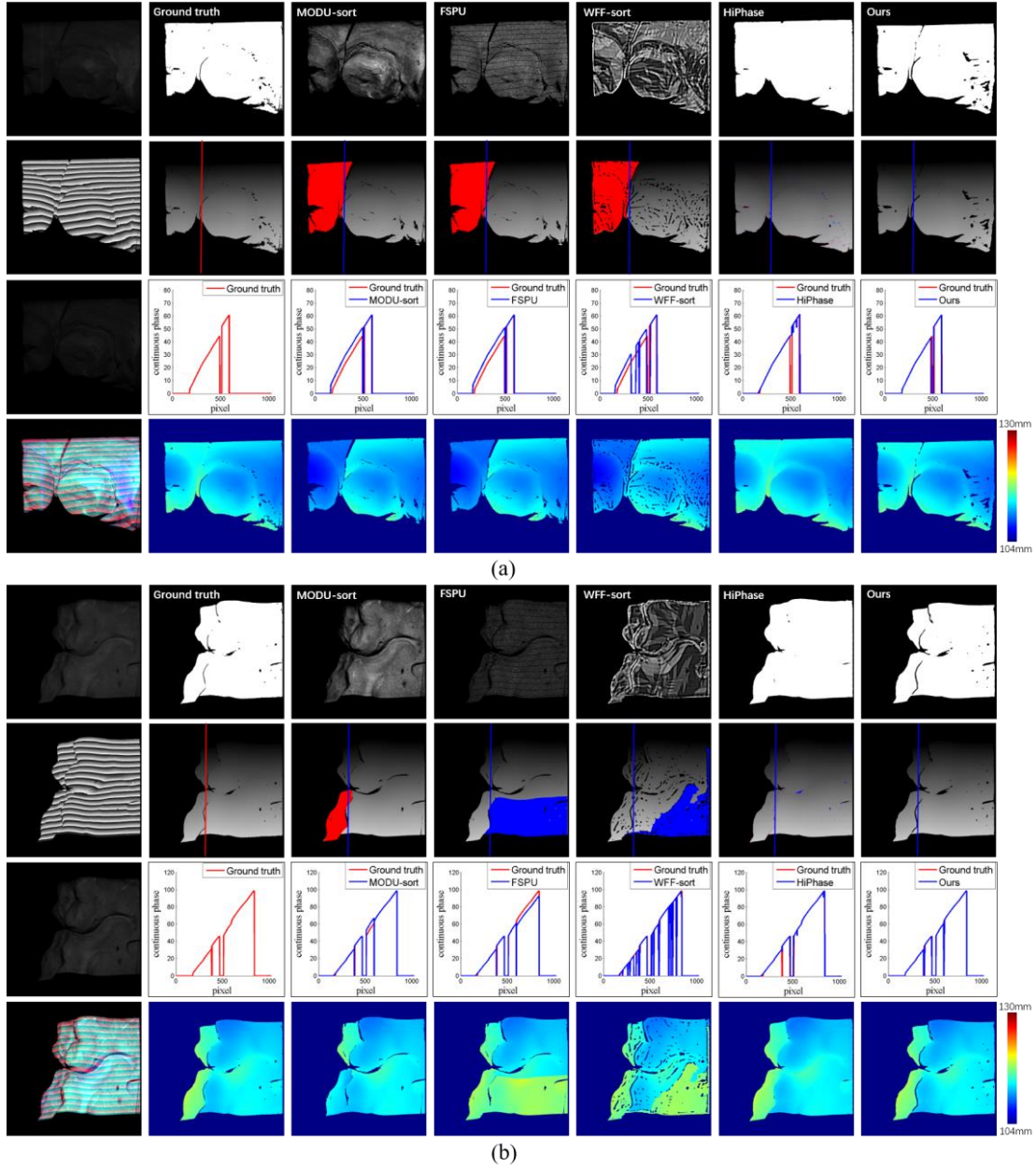

Fig. 15. Unwrapping results of five methods in phase discontinuity scenes. In each group, the first column shows the background intensity, phase map, intensity modulation, and PMI image from top to bottom. The second column is the ground truth. Results of the five methods are shown in the third to the seventh column; from top to bottom, the quality-map/unwrapping mask, continuous phase, phase profile, and depth map are illustrated. Red points in the continuous phase map indicate the points whose continuous phase is 0.3 rad bigger than their ground truth, and blue points show the points whose continuous phase is 0.3 rad lower than their ground truth.

There are 3,091 complex phase maps in Test-Data-A. Complex phase maps contain more than one nonideal type of surface reflectivity, motion blur, and phase discontinuity. For complex phase maps, finding the viable unwrapping path for QGSPU becomes more challenging. Under this condition, the number of failure cases of MODU-sort, FSPU, WFF-sort, HiPhase, and our method is 1,143, 1,567, 1,323, 386, and 180, respectively. Figure 16 depicts two examples of unwrapping results of the five methods in complex scenarios.

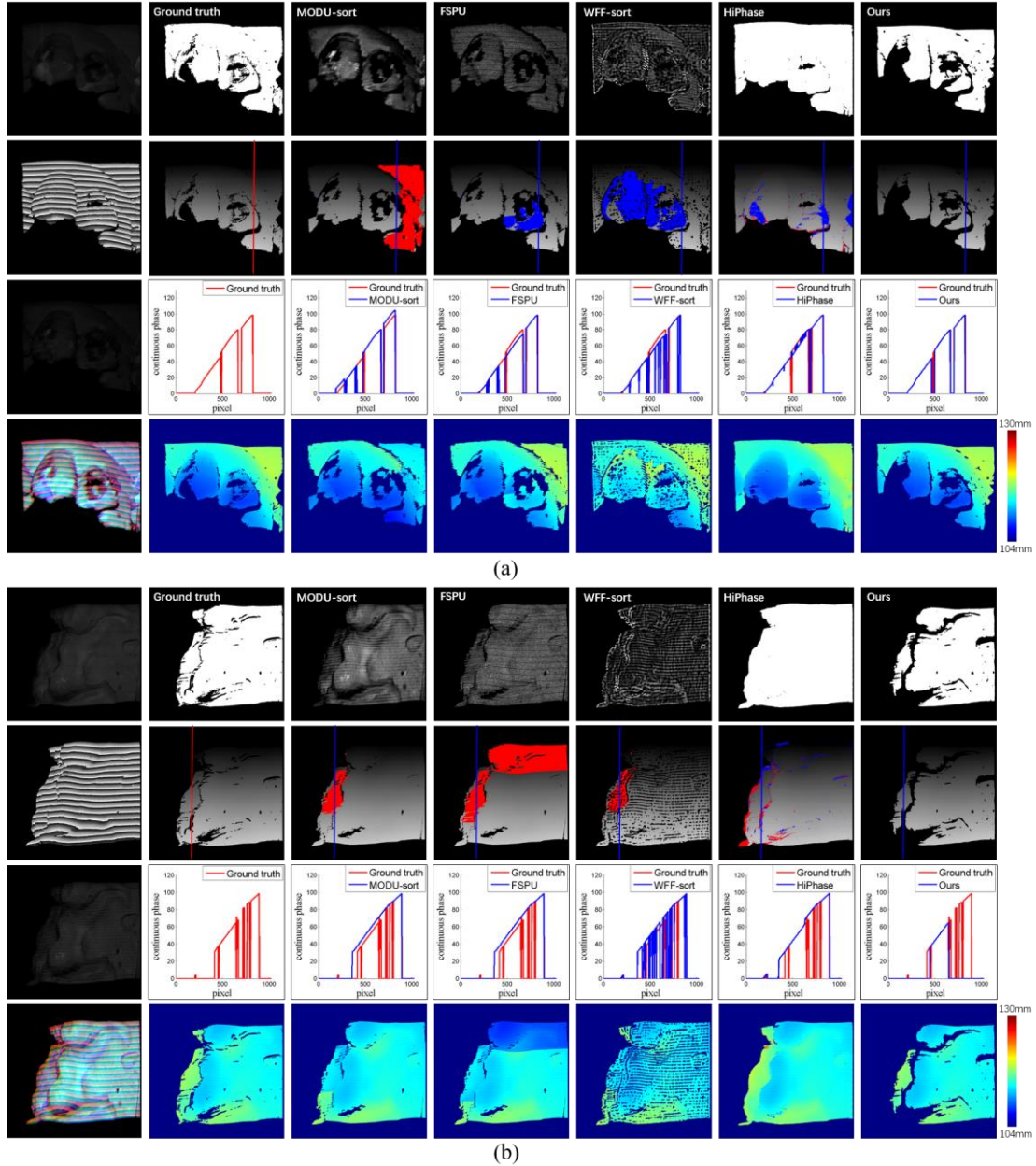

Fig. 16. Unwrapping results of five methods in complex scenarios. In each group, the first column shows the background intensity, phase map, intensity modulation, and PMI image from top to bottom. The second column is the ground truth. Results of the five methods are shown in the third to the seventh column; from top to bottom, the quality-map/unwrapping-mask, continuous phase, phase profile, and depth map are illustrated. Red and blue points in the continuous phase indicate inconsistent points.

*3.6 3D reconstruction using our method*

Successive scans of fringe images were used to test the effectiveness of our method in handheld FPP applications. The relative-to-absolute fringe order adjustment is accomplished using the method by [30], which projects an additional pattern during each scan. This additional pattern only contains one bright line of known absolute fringe order. Through triangulation, a point cloud is derived from the recovered continuous phase. All point clouds are registered and fused into one complete 3D model. During the process of continuous 3D reconstruction, none of the phase maps were manually excluded. Figure 17 illustrates the 3D reconstruction of one gypsum statute unseen by our network previously.

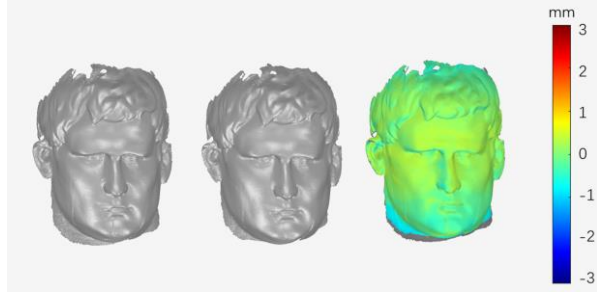

Fig. 17. Continuous 3D reconstruction of one unseen object with our method. There are 1,290 frames of phase maps that were unwrapped and transformed into depth maps for continuous registration and fusion. The left is the 3D reconstruction from the continuous phase of the TPU method. The middle is from the continuous phase of our methods. The right is the deviation plot.

*3.7 Generality test of our method*

The generality of the trained neural network on different illuminations of the same FPP system was demonstrated in Section 3.5. We further verify the generality of the trained neural network on other FPP systems using the dataset of Test-Data-B, Test-Data-C, and Test-Data-D. Note that these data are significantly different from the training dataset in object shape, optics wavelength, measurement area, number of fringes, and image contrast. And the image resolution and fringe direction of Test-data-B and Test-data-C are different from those of the training dataset. The ground-truth continuous phase of all test datasets is recovered by TPU algorithms and filtered with a modulation threshold of two to remove background points and small regions removal.

Figures 18-20 show our method's unwrapping results on Test-data-B, Test-data-C, and Test-data-D. Although our network did not see any data from these different FPP systems, the unwrapping results show that it has good transfer capability on these zero-shot data.

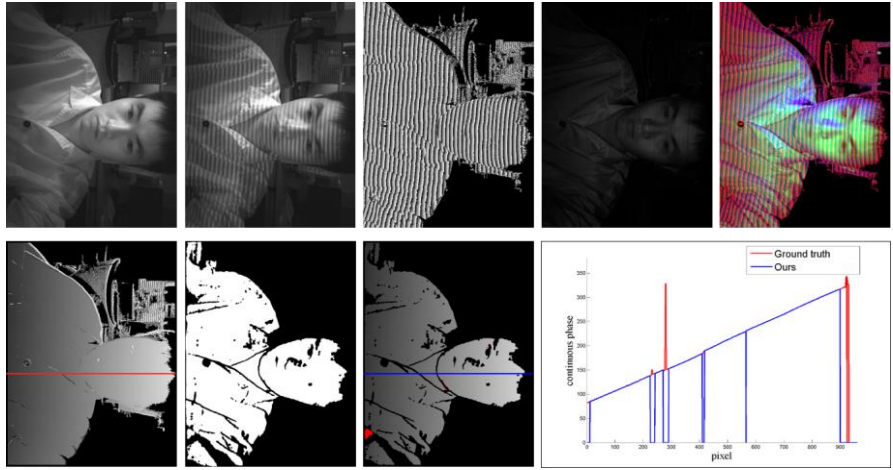

Fig. 18. Unwrapping results of our method on Test-data-B collected on an optics wavelength of 820nm. The first row is the background intensity, fringe image, wrapped phase, intensity modulation, and PMI image. The second row shows the ground-truth continuous phase, the unwrapping mask and continuous phase from our method, and the phase profile. Red points in the continuous phase indicate inconsistent points. The resolution of fringe images is 960×1280 pixels. The continuous phase recovered with our method is manually aligned with the ground truth.

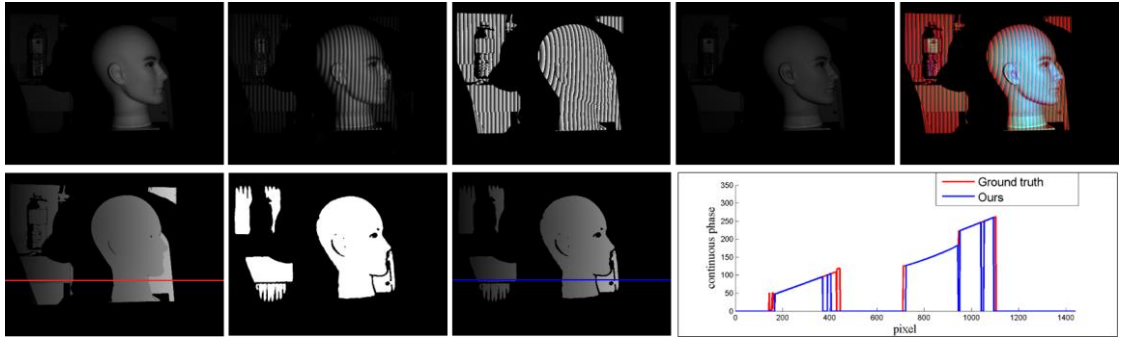

Fig. 19. Unwrapping results of our method on Test-data-C. This data is collected on an optics wavelength of 730nm. The first row shows the background intensity, fringe image, wrapped phase, intensity modulation, and PMI image from left to right. The second row shows the ground-truth continuous phase, the unwrapping mask, the continuous phase recovered with our method, and the phase profile. The resolution of fringe images is 1440×1080 pixels. The continuous phase recovered with our method is manually aligned with the ground truth.

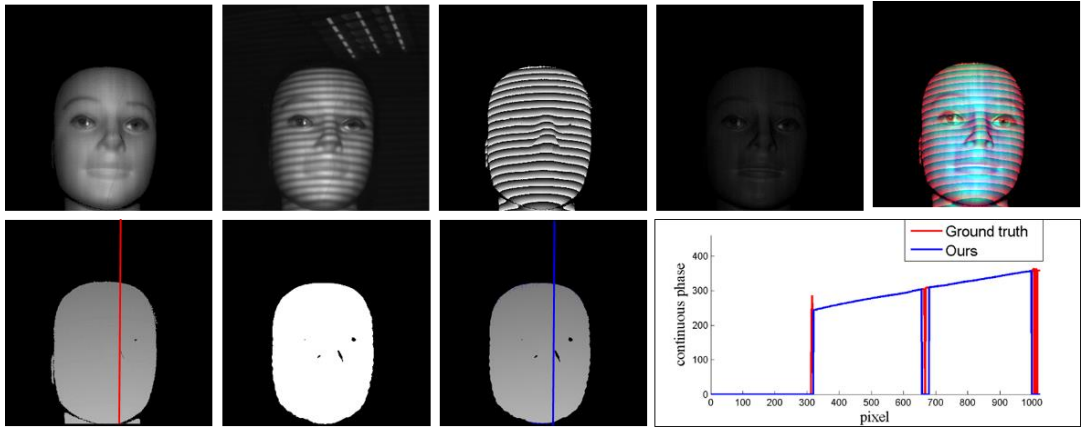

Fig. 20. Unwrapping results of our method on Test-Data-D collected with one white-light DLP FPP system. The first row shows the fringe image, wrapped phase, intensity modulation, background intensity, and PMI image from left to right. The second row shows the ground-truth continuous phase, the unwrapping mask, the continuous phase recovered with our method, and the phase profile. The continuous phase recovered with our method is manually aligned with the ground truth.

## 4. Discussion

### 4.1 Effect of error-percent threshold on comparison results

The error-percent threshold that determines whether one phase map is one failure case or not impacts the comparison results between different methods. This threshold is set at 0.1% in the experiments mentioned above. This means that if the number of points in an inconsistent subregion exceeds 0.1% of the total number of points of the phase map, the unwrapping is considered to have failed. To evaluate the effect of the error-percent threshold setting on the percentage of failure cases, we increase the error-percent threshold from 0.1% to 1% and count the percentage of failure cases of different methods on Test-data-A. Figure 21 shows that the percentage of all methods' failure cases drops accordingly with the increase of the error-percent threshold. At a threshold of 1%, our method achieves a failure case percentage of 2.40%, which is significantly better than the other four methods.

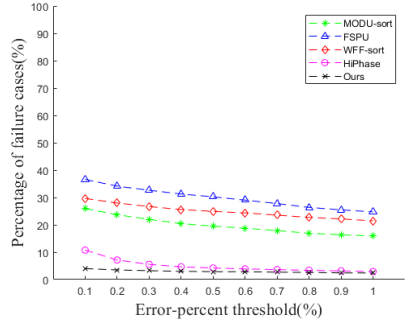

Fig. 21. Failure case percentage of the five methods at different error-percent thresholds on Test-data-A.

*4.2 Effect of the masking threshold on the results of QGSPU methods*

For QGSPU methods, increasing the masking threshold will reduce the number of failure cases in complex scenes. Among all the 3,091 complex phase maps in Test-Data-A, our method can correctly unwrap 2,911 maps. We count the average percentage of remaining points after masking before unwrapping, as shown in Fig. 8. The three QGSPU methods remove more points than our method before unwrapping, but the three QGSPU methods are inferior to our method.

*4.3 Advantages of our method*

This study aimed to improve the performance of the SPU method for FPP in real scenarios with deep learning. The proposed method has the following advantages,

- **Better performance than main QGSPU methods**. Previous research by Zhao et al. [16] compared the performance of different QGSPU methods in the presence of noise and phase discontinuity. It found that none of the quality maps could best guide the phase unwrapping process for phase discontinuity. By virtue of the advantages of deep learning in extracting image features and effectively integrating the information of different maps, the proposed method effectively detects unreliable points in complex phase maps. In section 3.5, experimental results show that the proposed method outperforms three main QGSPU methods in low surface reflectivity, motion blur, phase discontinuity, and complex scenarios. On our test dataset (i.e., Test-data-A) containing 5,632 phase maps acquired under five illuminations, at the error-percent threshold of 0.1%, the failure case percentage of our method is 4.01%; in contrast, the failure case percentage of MODU-sort, FSPU, WFF-sort, and Hi-Phase is 25.98%, 36.59%, 29.62%, and 10.76%, respectively. When the error-percent threshold is set at 1%, the failure case percentage of our method drops to 2.40%; in contrast, the failure percentage of MODU-sort, FSPU, WFF-sort, and Hi-Phase drop to 15.89%, 24.73%, 21.47%, and 2.95%, respectively. For many applications, it is acceptable that the number of points with unwrapping error in a phase map does not exceed 1% of the total number of pixels.

- **Avoid manual adjustment of the quality map threshold.** In applications, setting a quality map threshold using QGSPU methods is indispensable to exclude some unreliable points from the unwrapping process. Whether this threshold is appropriate or not has a significant impact on the quality of the unwrapped phase. The ideal threshold is related to the scene illumination, configuration of camera and projection unit, and surface reflectivity of the measured object. For better performance, manual adjustment of this threshold is usual. The proposed method avoids the manual adjustment of the quality map threshold. The pre-trained network can detect unreliable points in phase maps of various scenes and FPP systems, even though these scenes or systems significantly differ in measurement range, image contrast, number of fringe patterns, and fringe direction. In contrast, although relative ideal thresholds (i.e., modulation threshold of 6, 7, 8, 9, and 10) were set for various illuminations, MODU-sort still failed to unwrap 25.98% phase maps of the Test-data-A.

- **Better interpretability than end-to-end deep learning SPU methods.** Although deep learning is used, unlike the end-to-end scheme, the proposed method adopts a two-stage hybrid scheme. Deep learning is employed to estimate a classification map, and the flood-fill algorithm is used to unwrap the phase map under the guidance of this classification map. In this classification map, each point is predicted as background points, unreliable object points, and reliable points. The pointwise classification map output from the neural network is intuitive and understandable, and it can be directly used to analyze the causes of unwrapping right or wrong.

- **Generality in never-before-experienced situations**. In terms of the generality of deep learning-related methods, which are generally concerned, the experiments demonstrate the generality capability of the proposed method on scenes of different illumination conditions and multiple cross-system data, although the neural network did not see them during the training. Multiple cross-system data are from three different FPP systems and vary in image contrast, the number of fringes, image resolution, and fringe direction. This further verifies the advantages of the hybrid SPU scheme over the end-to-end scheme.

*4.4 Limitations*

The proposed method also has some limitations. First, out of all the 5,632 phase maps in Test-data-A, 4.01% (at an error-percent threshold of 0.1%) or 2.40% (at an error-percent threshold of 1%) of phase maps the proposed method failed to unwrap. The primary reason for these failure cases is that the proposed method still works on the assumption of local smoothness. The second limitation is that holes are left in the continuous phase after the unreliable points are removed from the phase maps. Actually, some points in these holes may be reliable. Better approaches to reliable point detection could reduce the size of the holes. Finally, a better network could accomplish a similar task to this study.

## 5. Conclusion

The present study aims to use deep learning to improve the performance of SPU for FPP in complex scenarios. A hybrid deep learning SPU scheme is proposed, in which a deep neural network is employed to learn the discriminative features of unreliable points from the composite images before unwrapping, and flood-fill is used to unwrap the filtered phase maps. Experimental results on a large dataset of real scenes demonstrated that this hybrid deep learning SPU scheme performs significantly better than conventional QGSPU methods. In addition, the proposed scheme avoids the manual adjustment of the masking threshold to adapt to the different scenes. The presented method has better interpretability than the end-to-end deep learning SPU method. The generality of the presented method in never-before-experienced situations is demonstrated using the data from different illuminations and three FPP systems differing in the measurement area, image contrast, the number of fringes, image resolution, and fringe direction, and optics wavelength.

**Funding.** This work was supported by Science and Technology Department of Sichuan Province (2022ZDZX0031) and Research Program (RD-03-202003) of West China Stomatology Hospital, Sichuan University.

**Acknowledgments.** Portions of this work were presented at the Proc. of SPIE, v12169, 2022, Eighth Symposium on Novel Photoelectronic Detection Technology and Applications in 2022, Learning-based invalid points detection for fringe projection profilometry.

**Disclosures.** The authors declare no conflicts of interest.

**Data availability.** The training dataset, source code, and pre-trained network will be available at https://github.com/WanzhongSong/DL-SPU.